\documentclass[preprint,12pt,sort&compress]{elsarticle}

\usepackage{amssymb}
\usepackage{mathtools}
\usepackage[T1]{fontenc}
\usepackage{listings}
\usepackage{hyperref}
\usepackage{slashed}
\usepackage{upquote}
\usepackage{mmacells}

\mmaDefineMathReplacement[≤]{<=}{\leq}
\mmaDefineMathReplacement[≥]{>=}{\geq}
\mmaDefineMathReplacement[≠]{!=}{\neq}
\mmaDefineMathReplacement[→]{->}{\to}[2]
\mmaDefineMathReplacement[⧴]{:>}{:\hspace{-.2em}\to}[2]
\mmaDefineMathReplacement{∉}{\notin}
\mmaDefineMathReplacement{∞}{\infty}

\mmaDefineMathReplacement{𝕕}{\mathbbm{d}}
\mmaSet{
  morefv={gobble=2},
  linklocaluri=mma/symbol/definition:#1,
  morecellgraphics={yoffset=1.9ex}
}

\newcommand{\fixlinespacing}{\linespread{1.18}}

\newcounter{bla}

\journal{Computer Physics Communications}

\newcommand{\FT}{{\emph{FormTracer}}}

\newcommand{\imag}{i}
\newcommand{\g}[1]{\gamma_{#1}}

\hypersetup{
	pdftitle={FormTracer - A Mathematica Tracing Package Using FORM},
	pdfauthor={A.~K. Cyrol, M. Mitter, N. Strodthoff},
	pdfkeywords={Trace} {FORM} {Mathematica} {Feynman diagrams},
	bookmarksopen=false,
	bookmarksnumbered=true
}

\def\Eq#1{Eq.~\eqref{#1}}
\def\Sec#1{Sec.~\ref{#1}}

\DeclareMathOperator{\Tr}{Tr}

\begin{document}

\begin{frontmatter}

\title{FormTracer\\A Mathematica Tracing Package Using FORM}

\author[a]{Anton K. Cyrol\corref{cyrol}}
\author[a]{Mario Mitter}
\author[b,a]{Nils Strodthoff}

\cortext[cyrol]{Corresponding author.\\\textit{E-mail address:} cyrol@thphys.uni-heidelberg.de}
\address[a]{Institut f\"ur Theoretische
  Physik, Universit\"at Heidelberg, Philosophenweg 16, 69120 Heidelberg, 
Germany}
\address[b]{Nuclear Science Division, Lawrence Berkeley National Laboratory, 
Berkeley, CA 94720, USA}

\begin{abstract}
We present \FT, a high-performance, general purpose, easy-to-use Mathematica 
tracing package which uses FORM. It supports arbitrary space and spinor 
dimensions as well as an arbitrary number of 
simple compact Lie groups. While keeping the usability of the Mathematica 
interface, it relies on the efficiency of FORM.
An additional performance gain is achieved by a decomposition algorithm that 
avoids redundant traces in the product tensors spaces.
\FT\ supports a wide range of syntaxes which endows 
it with a high flexibility. Mathematica notebooks that automatically install 
the package and guide the user through performing standard traces in 
space-time, spinor and gauge-group spaces are provided.
\end{abstract}


\end{frontmatter}

{\bf PROGRAM SUMMARY}

\begin{small}
\noindent
{\em Program Title:}  FormTracer                                          \\
{\em Licensing provisions:} GPLv3                                   \\
{\em Programming language:} Mathematica and FORM                                  \\
{\em Computer:}  any computer running FORM 4.1 and Mathematica 10 or higher                            \\
{\em Nature of problem:}  efficiently compute traces of large expressions\\
{\em Solution method:}\\
The expression to be traced is decomposed into its subspaces by a recursive 
Mathematica expansion algorithm. The result is subsequently translated to a FORM 
script that takes the traces. After FORM is executed, the final 
result is either imported into Mathematica or exported as 
optimized C/C++/Fortran code.\\
{\em Unusual features:}\\
The outstanding features of \FT\ are the simple interface, the capability to 
efficiently handle an arbitrary number of Lie groups in addition to Dirac and 
Lorentz tensors, and a customizable input-syntax.
   \\

\end{small}

\section{Introduction}
\label{sec:introduction}
Computer-algebraic tools are the backbone of many recent advances in 
theoretical high energy physics. This is particularly true in perturbation 
theory, where a large variety of tools for different stages of these very 
demanding calculations is available, see 
\cite{Harlander:1998dq,Baur:2007ub,Luisoni:2016xkv} for 
reviews. But also recent advances in the application of non-perturbative 
continuum functional methods, see 
\cite{Berges:2000ew,Roberts:2000aa,Alkofer:2000wg,Pawlowski:2005xe,
Fischer:2006ub,Gies:2006wv,Schaefer:2006sr,Binosi:2009qm,
Braun:2011pp,Maas:2011se,Eichmann:2016yit} for reviews,
require the ability to perform traces over increasingly complicated algebraic 
expressions. These stem mainly from the need to evaluate the one- and two-loop diagrams of 
these methods at off-shell momenta as well as expressing the appearing products of dressed 
propagators and vertices in terms of ever larger bases of tensors structures, see e.g.\ 
\cite{Windisch:2012de,Eichmann:2014xya,Williams:2014iea,Gracey:2014mpa,
Cyrol:2014kca,Mitter:2014wpa,Williams:2015cvx,Huber:2016tvc,Cyrol:2016tym}.
To summarize, typical workflows in perturbative as well as non-perturbative 
calculations involve the evaluation of traces in different subspaces, ranging 
from Lorentz and Dirac to group traces.

Here, we present \FT, a dedicated tracing tool that uses 
FORM \cite{Vermaseren:2000nd:elsevier,Kuipers:2012rf} in connection with an efficient 
expansion algorithm in Mathematica. Applying \FT\ requires only basic 
Mathematica skills, since all its features can be accessed directly from the 
Mathematica interface. The translation into FORM scripts is completely 
automated and consequently no knowledge of FORM syntax is required.
The final results are either imported into Mathematica for further 
manipulations or exported as optimized C/C++/Fortran code with FORM's optimization 
routine \cite{Kuipers:2013pba}. Furthermore, many examples and an automatic 
installation routine are provided in Mathematica notebooks that include all the 
necessary information for performing traces in standard applications.

Crucial parts of \FT\ have been developed during the course of 
\cite{Cyrol:2014kca,Mitter:2014wpa}. By now, \FT\ represents an integral part of 
the workflow within the fQCD collaboration \cite{fQCD:2016-10} and has been used during 
the derivation of the equations in 
\cite{Mitter:2014wpa,Braun:2014ata,Rennecke:2015eba,Eichhorn:2016esv,Cyrol:2016tym}.

Since most of the information needed for applying \FT\ is contained already in 
the Mathematica notebooks referred in \Sec{sec:installation}, the 
remainder of this paper focuses on providing additional theoretical 
background. In particular, we give a list of features in \Sec{sec:features} 
and summarize relevant properties of simple compact Lie groups in 
\Sec{sec:groups}. 
The treatment of the fifth gamma matrix in general dimensions as well as finite 
temperature and density applications are discussed in \Sec{sec:LorentzDirac}. 
We compare \FT\ to similar tools in \Sec{sec:others} and
explain the decomposition algorithm in \Sec{sec:decomposition}.
We discuss further algorithmic details in \Sec{sec:further_algorithms}
before we conclude in \Sec{sec:conclusion}.

\section{Installation and Usage}
\label{sec:installation}
\subsection{Installation and Quickstart Guide}
\fixlinespacing
\FT\ requires FORM\footnote{
FORM is licensed under the GPLv3.
The source code can be obtained from \cite{github:form} and
readily compiled executables are available at \cite{form:homepage}.
} 
version 4.1 and Mathematica 10.0 or higher.
We recommend to install \FT\ with the fully automated installation script,
which can be downloaded and started by evaluating
{
\footnotesize
\begin{mmaCell}{Code}
  \mmaDef{Import}["https://raw.githubusercontent.com/FormTracer/\
	 FormTracer/master/src/FormTracerInstaller.m"]
\end{mmaCell}
\par
}
\noindent in a Mathematica input cell. If FORM is not already installed on your computer,
it can be installed automatically during the installation process.
As an alternative to the automatic installation, one can also download \FT\ manually from \cite{github:FormTracer}
and install it by copying it into Mathematica's application folder.
Three notebooks with examples are available for download:
\begin{itemize}
	\item https://raw.githubusercontent.com/FormTracer/FormTracer/\\
	\href{https://raw.githubusercontent.com/FormTracer/FormTracer/master/src/Examples/FormTracerShowcase.nb}{master/src/Examples/FormTracerShowcase.nb}
	\item https://raw.githubusercontent.com/FormTracer/FormTracer/\\
	\href{https://raw.githubusercontent.com/FormTracer/FormTracer/master/src/Examples/FormTracerMinimalExample.nb}{master/src/Examples/FormTracerMinimalExample.nb}
	\item https://raw.githubusercontent.com/FormTracer/FormTracer/\\
	\href{https://raw.githubusercontent.com/FormTracer/FormTracer/master/src/Examples/FourQuarkInteraction.nb}{master/src/Examples/FourQuarkInteraction.nb}
\end{itemize}
The first notebook provides an extensive overview over
the features of \FT\ based on many example traces, whereas the 
second
provides the minimal prerequisite for being able to perform simple traces over 
space-time, spinor and gauge group indices.
To demonstrate \FT's performance, we provide a third notebook with a more complicated
example, namely the tracing of four-quark interaction diagrams performed in \cite{Mitter:2014wpa}.
All examples can also be found in the examples folder in the installation directory. The first two
examples provide an input cell to execute the automatic installation script as shown above.
While these example files
should be understood as quickstart guides to start using \FT\
as fast as possible, we also provide an extensive documentation 
in Mathematica's Documentation Center. A good overview
is given on the main page that can be accessed by simply searching for
\mbox{\emph{FormTracer}} in the Documentation Center.

\subsection{Basic Usage Examples}
\fixlinespacing
Once installed, \FT\ is loaded via
{
\footnotesize
\begin{mmaCell}{Code}
  \mmaDef{Needs}["FormTracer`"]
\end{mmaCell}
\par
}
\noindent \FT\, requires to define a custom notation which makes it easily adaptable
to the output of external diagram generators. Below, we define our notation for Lorentz tensors and
group tensors for two $SU(N)$ groups. For more information on the individual functions,
see the respective help pages in Mathematica's Documentation Center.
{
\footnotesize
\begin{mmaCell}{Code}
  \mmaDef{DefineLorentzTensors}[deltaLorentz[mu, nu], vec[p, mu], sp[p, q], 
      eps[], deltaDirac[i, j], gamma[mu, i, j], gamma5[i, j]];
  \mmaDef{DefineGroupTensors}[{
      {SUNfund, {color, Nc}, deltaAdj[a, b], 
          f[a, b, c], deltaFund[i, j], T[a, i, j]},
      {SUNfund, {flavor, Nf}, deltaAdjFlav[a, b], fFlav[a, b, c], 
          deltaFundFlav[i, j], TFlav[a, i, j]}}];
  \mmaDef{DefineExtraVars}[alpha, Mpsi, Zpsi, xi, g];
\end{mmaCell}
\par
}
\noindent \FT\ requires all further external variables to be declared before usage since
FORM requires it.
In the last line, we defined all variables that are used in the
examples below. Now, one can start tracing
{
\footnotesize
\begin{mmaCell}{Code}
  \mmaDef{FormTrace}[vec[p, nu] deltaLorentz[mu, nu] 
      vec[q, mu] deltaFund[i, j] deltaFund[j, i]]
\end{mmaCell}
\begin{mmaCell}{Output}
  Nc sp[p, q]
\end{mmaCell}
\par
}
\noindent or
{
\footnotesize
\begin{mmaCell}{Code}
  \mmaDef{FormTrace}[vec[p,nu]vec[q,rho]gamma[mu, i1,i2]
      gamma[nu,i2,i3]gamma[rho,i3,i4]gamma[mu,i4,i1]]
\end{mmaCell}
\begin{mmaCell}{Output}
  16 sp[p, q]
\end{mmaCell}
\par
}
\noindent \FT\ supports a shorthand notation for Dirac matrices that allows to 
leave out auxiliary indices. For example, one can evaluate
the above Dirac trace $\Tr\,\gamma^\mu \slashed p \slashed q \gamma^\mu$ 
simply via
{
\footnotesize
\begin{mmaCell}{Code}
  \mmaDef{FormTrace}[gamma[{mu,vec[p],vec[q],mu}]]
\end{mmaCell}
\begin{mmaCell}{Output}
  16 sp[p, q]
\end{mmaCell}
\par
}
\noindent As a more complex example from QCD, consider one-loop quark contribution
to the gluon propagator that is given by the following uncontracted expression
{
\footnotesize
\begin{mmaCell}{Code}
  testExpr = deltaAdj[colAdja, colAdjb]*
   (deltaLorentz[Mu, Nu] + xi (vec[p, Mu] vec[p, Nu])/sp[p, p])*
   g*gamma[Mu, i1, i4]*deltaFundFlav[flavFunda, flavFundd]* 
   T[colAdja, colFunda, colFundd]*(deltaDirac[i2, i1] Mpsi + 
   I gamma[Rho, i2, i1] vec[-p - q, Rho] Zpsi )/
   (Mpsi^2 + sp[p + q, p + q] Zpsi^2)* 
   deltaFundFlav[flavFundb, flavFunda]*
   deltaFund[colFundb, colFunda]*g*gamma[Nu, i3, i2]*
   deltaFundFlav[flavFundc, flavFundb]*
   T[colAdjb, colFundc, colFundb]*
   (deltaDirac[i4, i3] Mpsi + I gamma[Sigma, i4, i3]* 
   vec[q, Sigma] Zpsi)/(Mpsi^2 + sp[q, q] Zpsi^2)*
   deltaFundFlav[flavFundd, flavFundc]*
   deltaFund[colFundd, colFundc];
\end{mmaCell}
\par
}
\noindent which yields
{
\footnotesize
\begin{mmaCell}{Code}
  Simplify[\mmaDef{FormTrace}[\mmaDef{testExpr}]]
\end{mmaCell}
\begin{mmaCell}{Output}
  (2 \mmaSup{g}{2} (-1 + \mmaSup{Nc}{2}) Nf (2 xi \mmaSup{Zpsi}{2} sp[p, q] sp[p, p + q] + 
     sp[p, p] (\mmaSup{Mpsi}{2} (4 + xi) - (2 + xi) \mmaSup{Zpsi}{2} sp[q, p + q])))/
     (sp[p, p] (\mmaSup{Mpsi}{2} + \mmaSup{Zpsi}{2} sp[q, q]) (\mmaSup{Mpsi}{2} + 
     \mmaSup{Zpsi}{2} sp[p + q, p + q]))
\end{mmaCell}
\par
}
\noindent For further usage examples consult the documentation in Mathematica's 
Documentation Center or one of the example notebooks.

\section{Features}
\label{sec:features}
Here, we summarize the main features of \FT:
\begin{itemize}
\item evaluation of (Euclidean) Lorentz/Dirac traces in arbitrary dimensions, 
and traces over an arbitrary number of group product spaces, see 
\Sec{sec:groups}
\item high performance due to FORM backend combined with an efficient 
decomposition algorithm in Mathematica, see \Sec{sec:decomposition}
\item supports
\begin{itemize}
	\item the $\g5$ matrix in general dimensions within the Larin scheme, see \Sec{sec:Dirac}
	\item a special time-like direction for (Euclidean) finite temperature and 
density applications, see \Sec{sec:Lorentz}
	\item partial traces involving open indices
	\item creation of optimized output (including bracketing) using FORM's 
optimization algorithm~\cite{Kuipers:2013pba} for further numerical processing 
in C/C++/Fortran
	\item user-defined combined Lorentz tensors and corresponding identities, e.g.\ 
(transverse and longitudinal) projectors and their orthogonality relations, for 
speedup
\end{itemize}
\item intuitive, easy-to-use and highly customizable Mathematica frontend
\item convenient installation and update procedure within Mathematica
\end{itemize}

\section{Simple Compact Lie Groups}
\label{sec:groups}
\FT\ includes different group tracing algorithms that are implemented in FORM.
The most general algorithm is provided by the FORM color package
\cite{vanRitbergen:1998pn} and allows to take traces of arbitrary simple 
compact Lie groups.
Furthermore, we include explicit tracing algorithms for the fundamental 
representation in $SU(N)$, $SO(N)$ and $Sp(N)$, adapted from routines published 
with the color package \cite{vanRitbergen:1998pn} that use the Cvitanovic 
algorithm \cite{Cvitanovic:1976am} with additional support for partial traces.
Finally, we include dedicated tracing algorithms for the fundamental 
representations in $SU(2)$ and $SU(3)$ that support partial traces, explicit 
numerical indices as well as transposed group generators.
The use of explicit numerical indices 
requires to work in explicit representations.
For $SU(2)$ and $SU(3)$ we choose generators proportional 
to Pauli and Gell-Mann matrices, respectively.
Note that the fundamental $SU(N)$ tracing algorithm also supports partial 
traces but does not guarantee the same degree of simplification as
the specific $SU(2)$ and $SU(3)$ routines.
Due to the modular structure of the tracing procedure, the 
inclusion of further tracing algorithms at a later stage is easily possible.

The definitions of the group constants follow those of the color package
\cite{vanRitbergen:1998pn}, which we repeat here for the reader's convenience. 
We consider simple Lie algebras
with (Hermitian) generators $T_R$, which obey
\begin{equation}
[T^a_R,T^b_R]=\imag f^{abc} T_R^c\,,
\end{equation}
where $f^{abc}$ denote the structure constants. 
The dimensions of the representation $R$ and the adjoint representation are 
denoted by $N_R$ and $N_A\,$, respectively.
Furthermore, we define quadratic Casimir operators $C_R$ and $C_A$ via
\begin{equation}
(T^a_R T^a_R)_{ij}=C_R \delta_{ij}\quad \text{and} \quad f^{acd}f^{bcd}=C_A \delta^{ab}\,.
\end{equation}
It only remains to fix the normalization:
\begin{equation}
\text{Tr}\; T_R^a T_R^b=I_2(R) \delta^{ab}\,,
\end{equation}
where $I_2(R)$ denotes the second-order index of the representation $R$. Note that all tracing algorithms except for the FORM color package produce tracing results just in terms of the dimensions $N_R$ and $N_A$
with all other group constants set to their default values. In the case of $SU(N)$ in the fundamental representation, these values are given by
\begin{align}\label{eq:group_invariant_values}
 C_R =\frac{N_R^2-1}{2 N_R} ,\quad C_A =N_R ,\quad I_2(R) =\frac{1}{2}\, .
\end{align}

\section{Dirac and Lorentz Tracing}
\label{sec:LorentzDirac}

\subsection{Dirac Traces in General Dimensions}
\label{sec:Dirac}
Although FORM has built-in support for Dirac Traces in $d$ dimensions it
does not come with a built-in solution for the handling of the fifth gamma matrix, which
is defined as an inherently four-dimensional object.
Nevertheless, the generalization of the $\g5$ matrix to $d$ dimensions is
very important, in particular for applications using dimensional regularization. 
The implementation of the
fifth gamma matrix in $d\neq 4$ dimensions represents a subtle procedure and 
different prescriptions exist.
Here, we closely follow \cite{Larin:1993tq,Moch:2015usa} and implement support for the 
fifth gamma matrix by means of the Larin scheme \cite{Larin:1993tq} translated to Euclidean space-time,
which exploits the relation
\begin{equation}
\g\mu\g5=\frac{1}{3!}\epsilon_{\mu\nu\rho\sigma}\g\nu\g\rho\g\sigma\,.
\label{eq:altDef}
\end{equation}
For non-trivial expressions containing $\g5$ matrices, \FT\ applies the following algorithm to every spin line:
\begin{enumerate}
	\item Replace any occurring $\g5\g5$ with the unit matrix, $\g5\g5 \rightarrow 1\,$.
	\item Read each Dirac subtrace such that no $\g5$ is found on the leftmost position and replace all $\g5$ matrices using \Eq{eq:altDef}.
	\item Contract all epsilon tensors that do not stem from step 2.
	\item Contract all remaining epsilon tensors.
	\item Perform Dirac trace in $d$ dimensions with FORM.
\end{enumerate}
The separate contraction of different sets of epsilon tensors in 3.\ and 4.\ is necessary, since
there is no Schouten identity for general $d\neq 4$, which guarantees the
equivalence of different contraction orders in four dimensions.
When only a single $\g5$ matrix needs to be traced, a faster procedure based on 
an implicit application of \Eq{eq:altDef} can be used \cite{Moch:2015usa}.
By setting \emph{FastGamma5Trace[True]}, \FT\ applies \Eq{eq:altDef}
to all but the last $\g5$ matrix, which is then traced with this faster strategy. 

Due to the intricacies of the definition of $\g5$ in $d\neq 4$ dimensions, 
we encourage users to ensure that the implemented prescription is 
suitable for their specific application. Particular caution is necessary in the
case of multiple disconnected spin lines in the presence of connecting epsilon tensors.

\subsection{Finite Temperature and Density Tracing}
\label{sec:Lorentz}
\FT\ has a built-in functionality for a special time-like direction that is
useful for Euclidean finite temperature and density applications. It supports the definition
of space-like vectors
\begin{align}
	\label{eq:finiteTvector}
p_s=\begin{pmatrix}0\\\vec{p}\end{pmatrix},
\end{align}
which hold the spatial components of the corresponding full vectors
\begin{align}
	\label{eq:finiteTvector2}
	p=\begin{pmatrix}p_0\\\vec{p}\end{pmatrix}.
\end{align}
By definition, these vectors obey the following relations:
\begin{align}
	\label{eq:finiteTproperties}
	p \cdot_s q &\equiv  \vec{p} \cdot \vec{q} =p_s\cdot q_s = p_s \cdot q_{\phantom{t}} \,,
\end{align}
where the space-like inner product $\cdot_s$ has been introduced, which is supported by \FT.
In the evaluation of traces, the spatial vectors, $p_s$, with full dimensions 
are kept until the trace is performed. The traced
expressions are then represented in terms of standard 
and space-like inner products using \Eq{eq:finiteTproperties}.
This implementation of finite temperature is limited 
to Euclidean signature, which is sufficient for 
finite temperature and density applications in equilibrium.

\section{Comparison with Other Programs}
\label{sec:others}
\FT\ was designed for the specific task of evaluating Lorentz, Dirac and group 
traces. Our focus in its development was on usability, performance and the ability 
to handle very large expressions. In a typical workflow, these expressions are 
provided by further external programs that generate diagrams.
One example for such a tool with particular relevance for calculations in
non-perturbative functional methods is DoFun \cite{Huber:2011qr}.
In perturbative applications, input e.g.\ from FeynArts \cite{Hahn:2000kx} or QGRAF \cite{Nogueira:1991ex} 
as popular Feynman diagram generators is feasible.

There is a large number of tools, which have at least a partial overlap with \FT\ in
their functionalities, see for example \cite{Harlander:1998dq,Baur:2007ub,Luisoni:2016xkv}
for reviews on computer-algebraic methods in perturbative applications. 
In the following, we provide a comparison to programs that are from our point
of view most straightforwardly adaptable to tracing applications in the context of
non-perturbative functional method calculations. Although these tools were designed 
with more general applications in mind, we restrict the following comparison to their 
tracing capabilities.

\begin{itemize}

\item FORM \cite{Vermaseren:2000nd:elsevier,Kuipers:2012rf,Kuipers:2013pba,vanRitbergen:1998pn}
is a dedicated tool for high-performance symbolic calculations. It 
is a standalone program that comes with its own specialized input language. 
It has a built-in capability of taking Lorentz and Dirac traces as well as group 
traces involving arbitrary simple Lie groups using the color package 
\cite{vanRitbergen:1998pn}. The primary focus of FORM lies in speed and the 
ability to handle even very large symbolic expressions. However, the usage of 
FORM poses a rather steep learning curve for the beginner.
\FT\ aims to overcome this limitation by combining a 
Mathematica frontend in combination with a specialized expansion algorithm in 
Mathematica while still making use of the computational power of FORM in the 
background. By construction, it is always possible to write native FORM code for 
a specific tracing application that is as fast as the code generated 
automatically by \FT, but the latter is for many applications the more 
convenient choice. 

FormLink \cite{Feng:2012tk:elsevier} provides a way of accessing FORM via Mathematica to 
execute FORM commands and imports results back into Mathematica. However, it 
still requires the user to write FORM code and is therefore very close to FORM 
itself in its usage.

\item FeynCalc \cite{Mertig:1990an,Shtabovenko:2016sxi,Shtabovenko:2016whf}
is a popular Mathematica package for the symbolic 
semi-automatic evaluation of Feynman
diagrams and allows in particular to evaluate Dirac and Lorentz traces in 
arbitrary dimensions as well as fundamental $SU(N)$ group traces. 
Unlike \FT, it includes a rich set of tools beyond the ability of taking 
traces such as tensor reduction algorithms for one-loop integrals that make it 
particularly suited for perturbative applications. 
On the other hand, \FT\ is a specialized tracing tool and the complexity of 
expressions it can handle as well as its performance are typically only limited by FORM itself. 
As a consequence, \FT\ is often more than an order of magnitude faster 
than FeynCalc in examples typically occurring in non-perturbative functional QCD 
calculations.

\item HEPMath \cite{Wiebusch:2014qba} is a Mathematica package that extends the functionality of Tracer \cite{Jamin:1991dp}
and provides algorithms for high energy physics computations and therefore includes support for Lorentz, Dirac and 
$SU(3)$ group traces. In its functionality, HEPMath is similar to FeynCalc, which it aims to surpass
in usability and flexibility without focus on performance. It is 
designed as a convenience tool with a more general scope than \FT\ however
with a significantly smaller functionality in the tracing capability itself.

\end{itemize}

Apart from this incomplete selection, we want to acknowledge 
dedicated tools for the evaluation of diagrams such as
Mincer \cite{Larin:1991fz}, 
CompHEP \cite{Pukhov:1999gg:elsevier},
DIANA \cite{Tentyukov:1999is}, 
FormCalc \cite{Hahn:1998yk}, 
MATAD \cite{Steinhauser:2000ry}, 
GRACE \cite{Belanger:2003sd},
SANCscope \cite{Andonov:2004hi}, 
MadLoop \cite{Hirschi:2011pa},
GOSAM \cite{Cullen:2014yla}, 
Package X \cite{Patel:2015tea,Patel:2016fam} and 
Forcer \cite{Ueda:2016yjm}
as well as computer algebra systems/packages for tensor algebra with tracing capabilities like
GiNaC \cite{Bauer:2000cp}
GAMMA \cite{Gran:2001yh:elsevier}, 
Cadabra \cite{Peeters:2006kp},  
SymPy \cite{sympy:homepage}, 
xAct \cite{Martin-Garcia:2007bqa,xAct:homepage} and 
Redberry \cite{Bolotin:2013qgr:elsevier}.

\section{Decomposition of Tensor Classes}
\label{sec:decomposition}
This section explains details of the expansion algorithm, which is hidden from the user. 
Let \(x\) be an untraced expression, of which the trace over \(n\) Lie groups
as well as Dirac and Lorentz space is to be taken. 
The straightforward way to carry out the trace in  \(x\) would be to 
fully expand \(x\) into a sum of simple products of tensors and repeatedly apply 
the appropriate tensor identities to the summands.
However, this strategy almost always entails multiple calculations of identical subtraces.
Since FORM fully expands all expressions, we decompose \(x\) into its subspaces in Mathematica by
bringing it into the form
\begin{align}
	\label{eq:DecompositionGroups}
	x=c_0 \sum_{i_0} C_{i_0} \sum_{i_1} C_{i_0 i_1} \ldots \sum_{i_n} C_{i_0 i_1\dots i_n} L_{i_0 i_1\dots i_n}\,.
\end{align}
Here, \(c_0\) represents a scalar prefactor, \(C_{i_0\ldots i_j}\) contains 
only tensors of the \(j\)-th group, 
\(L_{i_0 i_1\dots i_n}\) consists only of Lorentz and Dirac tensors
 and the summation boundary of \(i_j\) in \Eq{eq:DecompositionGroups} 
depends on \(i_0,\ldots ,i_{j-1}\).
In addition to the considerable performance gain due to the uniqueness of the tensors \(C_{i_0\ldots 
i_j}\), this decomposition allows to take the traces of the individual Lie groups separately.
For tracing the combined Lorentz and Dirac tensors \(L_{i_0 i_1\dots i_n}\) in \Eq{eq:DecompositionGroups}, we provide two possibilities. By default,
no further manipulation is performed and we let FORM handle the evaluation of the \(L_{i_0 i_1\dots i_n}\)'s.

\section{Additional Algorithmic Details and Optimization}
\label{sec:further_algorithms}

\subsection{Partial Traces over Lorentz Tensors}
\label{sec:disentangle_lorentz_structures}
In some cases, where the scalars $L_{i_0 i_1\dots i_n}$ in \Eq{eq:DecompositionGroups} are very large,
a further decomposition can improve the performance by splitting the full trace into partial intermediate traces.
The decomposition can be turned on with the option \emph{DisentangleLorentzStructures[True]}, which 
first groups the expressions \(L_{i_0 i_1\dots i_n}\) into a sum of Lorentz and Dirac scalars
\begin{align}
	\label{eq:DecompositionLorentz}
	L_{i_0 i_1\dots i_n}= \sum_{j}\, l_{i_0 i_1\dots i_n; j}\, .
\end{align}
Next, each of these summands is written as,
\begin{align}
	\label{eq:DecompositionLorentz2}
	l_{i_0 i_1\dots i_n; j} &= \prod_{f=1}^{N_j}\left(\prod_{g=1}^{N_{jf}}l_{i_0 i_1\dots i_n; j f g}^{\{\mu_{jfg}\}}\right)\, .
\end{align}
Here, the factors of the outer product are the smallest possible Lorentz scalars and the factors of the inner product the smallest
possible, already traced, Dirac scalars that allow for such a representation. The superscripts ${\{\mu_{jfg}\}}$ denote a set of Lorentz indices 
and the factors are ordered such that 
\begin{align}\label{eq:DecompositionLorentz_ordering}
 \{\mu_{jfg}\}\cap \{\mu_{jf(g+1)}\} \neq \emptyset\ .
\end{align}
To avoid large expressions by possibly exploiting intermediate simplifications, the Lorentz traces are evaluated successively, 
\begin{align}
	\Tr \Bigg(\prod_{g=1}^{N_{jf}} & l_{i_0 i_1\dots i_n; j f g}^{\{\mu_{jfg}\}}\Bigg)\\
	&=\Tr  \left(l_{i_0 i_1\dots i_n;jf1}^{\{\mu_{jf1}\}} \Tr\left(\ldots 
	 \Tr  \left(l_{i_0 i_1\dots i_n;jf(N_{jf}-1)}^{\{\mu_{jf(N_{jf}-1)}\}} l_{i_0 i_1\dots i_n;jfN_{jf}}^{\{\mu_{jfN_{jf}}\}}\right)\right)\right)\,,\nonumber
\end{align}
and the results are multiplied and summed only at the very end.
This feature has been crucial for quantum gravity applications, in particular for tracing the four-graviton vertex equation \cite{Denz:2016qks:elsevier}.

\subsection{Other Algorithmic Improvements}
\label{sec:alg_improvements}
We implemented two further improvements that greatly reduce the number of terms in many of our standard applications.
Internal loop momenta are in general sums of the loop and external momenta.
Simply expanding all vector sums (e.g.\ \((p+q)_\mu\rightarrow p_\mu + q_\mu\)) leads 
to an unnecessarily large number of terms.
Thus, we replace sums of vectors by abbreviations and reinsert the explicit momenta after the tracing process.
Although this can prevent cancellations in the final result, we found these abbreviations to be very advantageous 
for the performance as well as the size of the final result in the general case involving off-shell momenta.

By default, FORM expands all powers of sums, even if the sums only contain scalars. In this special case, however, an expansion
is not necessary to evaluate the traces. Therefore, we prevent FORM from expanding powers over sums of scalars by using a 
user-defined power function symbol in the FORM code.

\section{Conclusion}
\label{sec:conclusion}
We presented the dedicated tracing package \FT\ for Mathematica. 
Its most notable features are its usability, performance, and the capability to efficiently handle an 
arbitrary number of Lie groups as well as Dirac and Lorentz tensors in arbitrary dimensions.
This includes an algorithm to deal with $\g5$ matrices in $d \neq 4$ dimensions.
\FT\ achieves its performance by using FORM as a powerful backend in combination with an advanced 
decomposition algorithm in Mathematica.
Furthermore, a simple but effective way to single out a special time-like direction for finite 
temperature and density applications in Euclidean field theory is provided.
Although developed with specific applications in non-perturbative functional methods in mind,
its flexible notation and usability facilitate the use in new and existing general purpose programs, 
in particular in perturbation theory.

\section*{Acknowledgement}
This work was done within the fQCD collaboration for which we particularly thank Jan M. Pawlowski. 
Furthermore, we are grateful to Andreas Rodigast and Jos Vermaseren for many valuable FORM tips and explanations
as well as to Vladyslav Shtabovenko for constructive comments especially with respect to $\g5$ matrices in general dimensions.
We acknowledge discussions with Nicolai Christiansen, Tobias Denz, Aaron Held, Jan M. Pawlowski, Manuel Reichert, and Nicolas Wink.

This work is supported by
the grant ERC-AdG-290623, 
FWF through Erwin-Schr\"odinger-Stipendium No.\ J3507-N27, 
the BMBF grant 05P12VHCTG, 
the Studienstiftung des deutschen Volkes, 
the DFG through grant STR 1462/1-1, and 
in part by the Office of Nuclear Physics in the US Department of Energy's Office of Science under Contract No. DE-AC02-05CH11231.


\bibliographystyle{elsarticle-num}
\bibliography{../../bib_master}

\end{document}